\documentclass[10pt,twocolumn]{article}
\usepackage{times}
\usepackage{cite}
\usepackage{amsmath,amssymb,amsfonts}
\usepackage{algorithmic}
\usepackage{graphicx}
\usepackage{textcomp}
\usepackage{xcolor}
\usepackage[normalem]{ulem}

\begin{document}

\title{An Optimal Resource Allocator of Elastic Training for Deep Learning Jobs on Cloud}
\author{
Liang Hu$^1$, Jiangcheng Zhu$^2$, Zirui Zhou$^1$, Ruiqing Cheng$^2$, Xiaolong Bai$^2$ and Yong Zhang$^1$ \\
\em  $^1$Huawei Technologies Canada \quad
          $^2$Huawei Cloud \\ [2mm]
\small{E-mail: lianghu814@gmail.com, \{zhujiangcheng, zirui.zhou, chengruiqing, baixiaolong1, yong.zhang3\}@huawei.com}
}

\date{}
\maketitle
\begin{abstract}
Cloud training platforms, such as Amazon Web Services and Huawei Cloud provide users with computational resources to train their deep learning jobs. Elastic training is a service embedded in cloud training platforms that dynamically scales up or down the resources allocated to a job. The core technique of an elastic training system is to best allocate limited resources among heterogeneous jobs in terms of shorter queueing delay and higher training efficiency. This paper presents an optimal resource allocator for elastic training system that leverages a mixed-integer programming (MIP) model to maximize the training progress of deep learning jobs. We take advantage of the real-world job data obtained from ModelArts, the deep learning training platform of Huawei Cloud and conduct simulation experiments to compare the optimal resource allocator with a greedy one as benchmark. Numerical results show that the proposed allocator can reduce queuing time by up to 32\% and accelerate training efficiency by up to 24\% relative to the greedy resource allocator, thereby greatly improving user experience with Huawei ModelArts and potentially enabling the realization of higher profits for the product. Also, the optimal resource allocator is fast in decision-making, taking merely 0.4 seconds on average.

\end{abstract}

\section{Introduction}

Cloud training platforms, such as Amazon Web Services and Microsoft Azure, provide abundant computational resources for training deep learning (DL) models and charge users by usage. Currently when users submit a deep learning job to the cloud, it is required to specify the desired computational resources, e.g., number of GPUs or nodes \cite{b1, b2, b3}. Cloud training platforms without elasticity will use the specified, also fixed, resources to train the job. However, There are at least two problems with using a fixed amount of resources for performance of the training job.

First, a system using a fixed number of nodes or GPUs for any given training job may use its computational resources inefficiently. If the system is performing a small number of training jobs at a given time, the system will leave many of its computational resources idle. In other words, the training jobs could have each utilized more nodes or GPUs in order to complete sooner, instead of wasting the computational capacity of the idle resources. For example, a system with 100 nodes performing only a single training job, wherein the fixed number of nodes assigned to the training job is 8 nodes, is wasting 92 of its nodes.

Second, the system’s computational resources are always limited by the size of the system’s resource pool, e.g., the number of nodes available for allocation to training jobs. It is common for a system to receive multiple job profiles from users while a small number of computationally-intensive training jobs are monopolizing the system’s resource pool, requiring the system to maintain the later job profiles in a job queue for a significant period of time while waiting for the computationally-intensive training jobs to complete. This introduces significant delays, even for small training jobs that could be completed quickly if any nodes were available. These delays are arguably inefficient in terms of meeting the needs of the user base, and tend to generate dissatisfaction in users who experience such delays.

\begin{figure}[h]
\centerline{\includegraphics[width=3.2in,keepaspectratio]{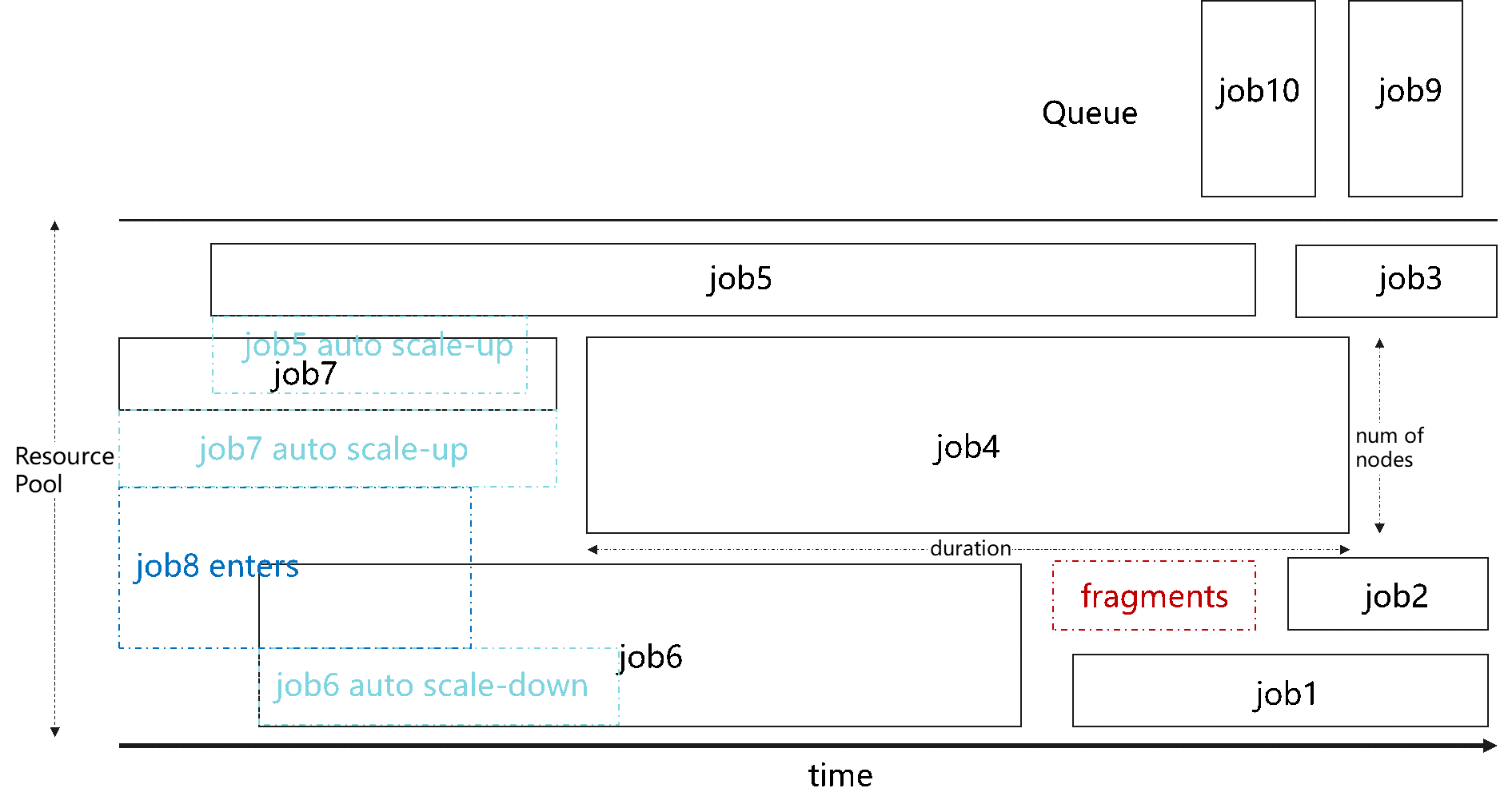}}
\caption{Resource pool and job queue of an elastic training system.}
\label{fig: resource pool}
\end{figure}

Accordingly, cloud systems have been developed that perform elastic training of deep learning models (referred to herein as “elastic training systems”) to address the limitations of systems using a fixed amount of resources for a given training job \cite{b4, b5}. An elastic training system dynamically allocates computational resources (e.g., nodes) to training jobs based on the status of the system (e.g., how many nodes are in use, how many jobs are in the job queue) and job attributes (e.g., how computationally intensive is a given training job) to address the two problems described above. If the system has abundant computational resources available (e.g., a large number of idle nodes), an elastic training system may scale up one or more ongoing training jobs, i.e., allocate more nodes or other computational resources to the one or more ongoing training jobs. If an elastic training system is busy (e.g., all nodes are being used for ongoing training jobs), the elastic training system scales down one or more of the ongoing jobs, i.e., releases some nodes or other computational resources so that new training jobs can use the released nodes or other computational resources instead of waiting in the job queue. Fig. \ref{fig: resource pool} is an example wherein different jobs are training on different number of nodes. If systems have no elastic training service, the resource pool may generate fragments indicating some nodes are being wasted. With elastic training service, job 5 and 7 may scale up to utilize more nodes, and job 6 may scale down to release part of its resources to job 8 that just enters the pool. 

The core of an elastic training system is its resource allocator. A resource allocator should optimally decide on the nodes or other computational resources assigned to each training job so that the elastic training system can (1) improve efficient utilization of computational resources, (2) speed up the overall training time required to complete a given set of training jobs, (3) reduce queueing delay, and (4) improve the user experience when submitting a job profile to the system. By achieving one or more of these objectives, and providing the benefits thereof to users, an elastic training system may also be able to realize higher profits in providing a paid deep learning product-as-a-service (PaaS) to users, through a combination of higher revenue from users due to improved service, and/or lower overhead costs due to more efficient use of resources.

Existing works include the mechanisms of elastic training and the resource management in an elastic training system. To achieve elasticity while training deep learning jobs, the industry has proposed software frameworks such as Google Kubernetes Engine (GKE) \cite{b6}, Amazon Elastic Container Service (ECS) \cite{b7}, and Red Hat OpenShift Container Platform \cite{b8}. Shen et al. \cite{b8.1} presented an automatic elastic resource scaling system, called CloudScale, for multi-tenant cloud computing infrastructures, though the infrastructures are not for the purpose of deep learning training specifically. Other than elasticity, resource management also plays a critical role in cloud system operations. One research \cite{b8.2} formulated the resource management problem for MapReduce jobs on the cloud as a constraint programming model. Its objective is to reduce energy consumption of MapReduce jobs. Liu and Xu \cite{b8.3} proposed an executor scheduler that is able to dynamically allocate and size resources to Spark jobs in order to minimize resource fragmentation. Another work by Javadi \textit{et al.} \cite{b8.4} also present a workload resource scheduler for Spark jobs that can significantly increase cloud resource usage.

In recent years studies on resource allocation for elastic training specifically are drawing more attention. Chen \textit{et al.} \cite{b8.5} found that the parameter server for deep neural networks training could become performance bottlenecks due to imbalanced workload distribution among the parameter servers. They designed a dynamic workload distribution scheme using an exploitation-exploration method in order to accelerate distributed model training. A similar study \cite{b8.6} also tried to solve the imbalanced workload distribution problem using a semi-dynamic load balancing approach, which accelerates distributed training by up to 54\%. Saxena \textit{et al.} \cite{b9} proposed a GPU-level resource allocator. The core idea is to find the best combination of batch size and number of GPUs to elastically train deep learning jobs. Their approach took efforts on maximizing total throughput of all jobs and searches for optimal combination through dynamic programming. However, this approach only works for GPU-level resource allocation, but not transferable to node-level. GPU-level resource allocation is a form of process management that manages the execution of individual software processes by individual processor cores. Resource allocation at the node level, however, implements a form of cluster management or container management, both of which refers to the management of containers (i.e., a bundle of a software program and its data dependencies) and their execution by virtualized clusters of processing resources. In addition, the allocate decisions in this paper do not make sure that the number of GPUs of a job is powers of two, which may lower training accuracy \cite{b10}. Parallel computing typically requires that computational operations to be split recursively by powers of two to avoid accuracy problems. Another practice in industry \cite{b10} tried to greedily utilize all computational resources at all times. Section \ref{section: greedy} thoroughly reviews this greedy resource allocator and we will use it as benchmark for comparison with our work.

This paper presents a novel resource allocator for elastic training systems that takes advantage of the ETA (estimated time of arrival), also known as estimated runtime, of deep learning jobs as input. We first formulate the resource allocation problem as a mixed-integer programming (MIP) model that maximizes the training progress of all jobs over a planning time horizon. The model also uses a innovative method to make sure that the allocated number of node to each job is $2^m$ ($m \in \mathbb{N}$). Obtaining the optimal resource allocate decisions is very fast. The decisions can significantly reduce queueing delay and improve training efficiency. In addition, the proposed allocator can better handle heterogeneous-ETA jobs and perform robustly on ETA disturbance and unexpected situations, such as jobs containing bugs and users terminating jobs by themselves.

This paper is organized as follows. Section \ref{section: architecture} introduces the  architecture of elastic training system of Huawei ModelArts. Section \ref{section: greedy} reviews a previous practice of Huawei ModelArts, i.e., the greedy resource allocator and its limitations. Section \ref{section: optimal} explains the methodology behind the optimal resource allocator and how we linearize and simplify the MIP model to make it solvable and efficient. Section \ref{section: simulation} takes advantages of real-world job data from Huawei ModelArts and conducts simulation experiments to compare the above two resource allocators. In the end Section \ref{section: conclusion} concludes the paper.

\section{Architecture of Elastic Training System of Huawei ModelArts}\label{section: architecture}

\begin{figure}[h]
\centerline{\includegraphics[width=3.2in,keepaspectratio]{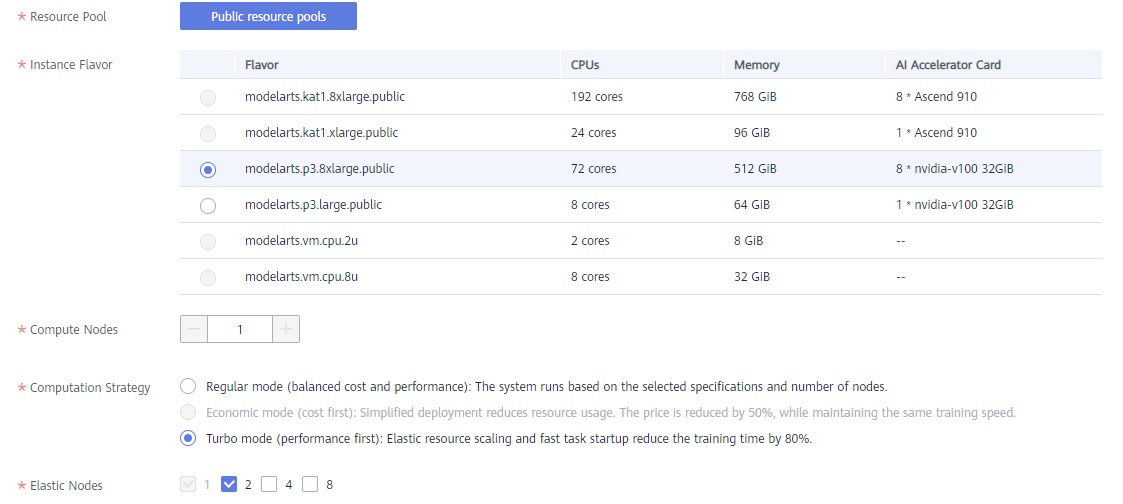}}
\caption{The elastic training service of Huawei ModelArts.}
\label{fig: modelarts ui}
\end{figure}

Huawei ModelArts is a one-stop artificial intelligence development platform of Huawei Cloud. Thousands of deep learning jobs are training on this platform every day. Elastic training is a service embedded in ModelArts that provides the capability to dynamically scale up or scale down distributive training jobs. Elastic training is critical to ModelArts since it holds the potential to improve computational resource utilization and accelerate user’s training jobs, and as a result, realize higher profits from users. Fig. \ref{fig: modelarts ui} shows a screenshot of ModelArts elastic training service. When a user submits a job profile to the platform, they need to specify the desired number of nodes required for performing the training job and the number of elastic nodes.

\begin{figure}[h]
\centerline{\includegraphics[width=3.2in,keepaspectratio]{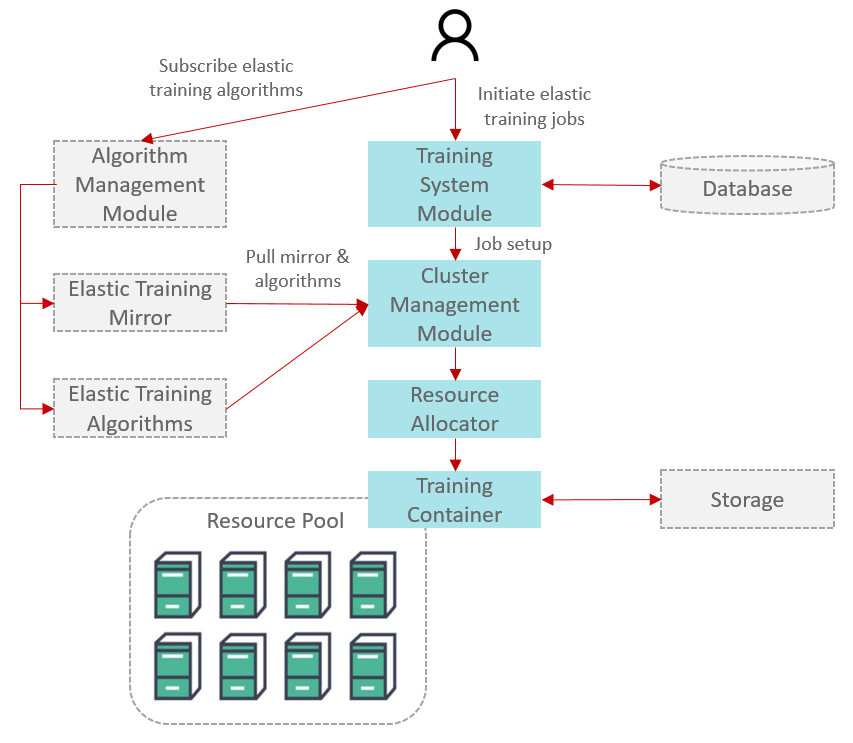}}
\caption{Architecture of elastic training system for deep learning jobs on cloud.}
\label{fig: patent system}
\end{figure}

The architecture of elastic training system of Huawei ModelArts consists of deep learning jobs, job queue, resource pool, resource allocator and other modules, as illustrated in Fig. \ref{fig: patent system}. Resource allocator is the core module and it acts as the ``brain'' of the system to make the best resource allocate decisions.

The job is defined as the workload to be run on certain computational resource, i.e., GPUs, and 1 node is made up of 8 GPUs. Jobs are trained distributively on different number of nodes. The resource allocator can make decision to scale-up or scale-down the computational resource of a certain job, i.e., number of nodes. The jobs adapt to the changes of its resources and keep on training without interruption. 

To initiate a training job, firstly users should publish an algorithm to the Algorithm Management Module, including the algorithm code and the algorithm mirror. The code and mirror will be pulled by the resource pool after the job is allocated to certain nodes by the resource allocator.

Once users have submitted a training job to the elastic training system, the training system transfers the application programming interface (API) body of the elastic training job into an API body that the cluster management module can address. The basic job profile information, including create time, starting time, flavor, etc. is stored in the database. The cluster management module monitors the status of each workload and each node, and makes real-time job allocation to the idle nodes. The computing node mounts the cloud storage, pulls the docker image from the hub, download the code, and then start the script.

The resource allocator is the core module of the system. It has access to all active jobs (queueing or training), job's remaining ETA, and total computational resources. Based on elastic training frequency $f$, e.g., 5 minutes, the resource allocator activates and makes decisions on scaling-up or scaling-down an active training job and how the nodes in the resource pool should be re-assigned. At the end, the decisions are sent back to job queue and resource pool for execution.

If some nodes are idle, the first job in the queue will start training using as many nodes as possible under a cap (e.g., 16 nodes). If idle nodes still exist, follow the same procedure to initiate the next queueing job until no queuing jobs or no idle nodes exist.


\section{Greedy Resource Allocator}\label{section: greedy}

One previous work \cite{b10} of Huawei Cloud proposed a greedy resource allocator for elastic training. This is a rule-based allocator that tries to greedily utilize as many nodes as possible. We use this allocator as benchmark for comparison with our work. Its rules are explained as follows.

The greedy resource allocator allocates the resource pool of the elastic training system based on four different scenarios, in which every training job is allocated a node count within a range, such as 1 to 16 nodes.

In the first scenario, the system has at least one idle node and at least one training job in the job queue. The greedy allocator allocates as many nodes as possible to the training job at the front of the job queue. If there are still idle nodes and training jobs in the job queue, this procedure is repeated until all nodes are occupied or all training jobs have exited the job queue and are being performed.

In the second scenario, the system has at least one idle node and no training jobs in the job queue. The greedy resource allocator finds the training job with the shortest training time, and then scales up this training job by increasing its node count as large as possible. If there are still idle nodes, this procedure is repeated until all nodes have been occupied or all training jobs have scaled up.

In the third scenario, the system has no idle nodes and at least one training job in the job queue. Thus, the computation resources of the system have reached their limit. Some training jobs might be occupying all the nodes while many others have to wait in the job queue. The greedy resource allocator finds the training job with the longest training time, scales down the training job through reducing its node count by half, and then allocates the released nodes to the training job at the front of the job queue.

In the fourth scenario, the system has no idle nodes and no training jobs in the job queue. This is the simplest scenario. All nodes are occupied and no training jobs are waiting. In this case, the elastic training system changes nothing about the current node allocation.

\begin{figure}[h]
\centerline{\includegraphics[width=3.2in,keepaspectratio]{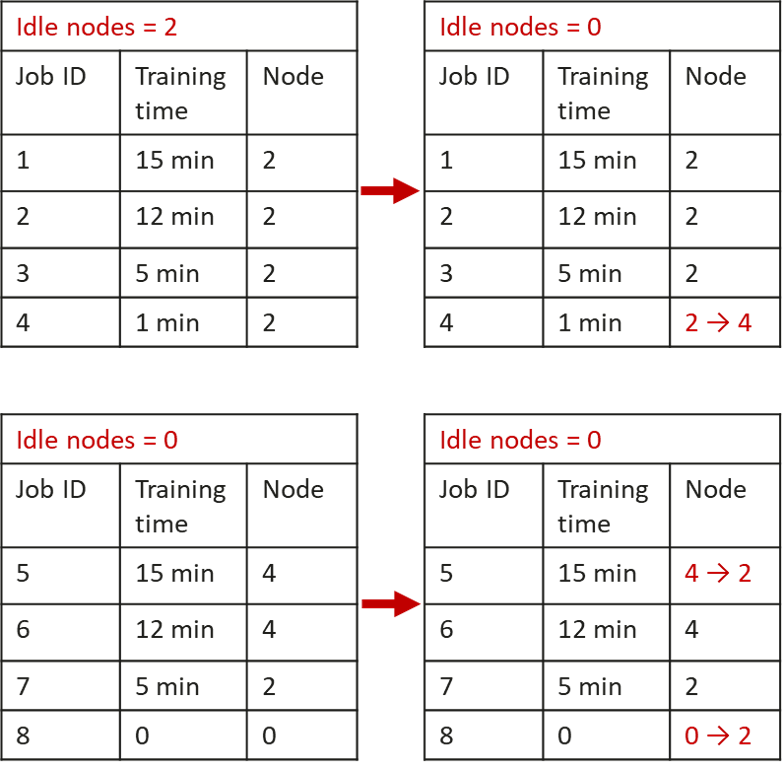}}
\caption{Two examples for greedy resource allocator.}
\label{fig: greedy}
\end{figure}

Fig. \ref{fig: greedy} gives two examples that help better understand the rules of greedy resource allocator. Assume the range of scaling is 1$\sim$16 nodes. For the first example on the top, the elastic training system has 2 idle nodes but no queueing jobs. There are four jobs training currently and job 4 has the shortest training time, so the 2 idle nodes are allocated to it. For the second example at the bottom, job 8 is waiting but the system has no idle nodes. Job 5 has the longest training time, so the allocator scales down its node size from 4 to 2 and assigns the released 2 nodes to job 8.

The resource allocator is called “greedy” because it always tries to utilize the system’s computational resources to their fullest extent, i.e., leave no nodes idle. The rules governing its behavior tend to be simple and computationally fast. However, the simplicity of its behavior results in several limitations.

First, while the greedy resource allocator keeps as many nodes working as possible, the allocation of nodes to training jobs may not be efficient or fair. For example, a greedy resource allocator may inefficiently allocate 99 nodes to job 1 and 1 node to job 2, instead of allocating 50 nodes to each job (job 1 and job 2). Although both allocations utilize all 100 nodes, the second one is obviously more equal and may result in more overall efficiency. 

Second, training time may not be a good metric to use in deciding which job should scale up or down. The greedy resource allocator scales up the job with the shortest training time, but if this job has a very small workload, one node may be sufficient; the additional nodes might be more effectively deployed to a larger training job. Similarly, the greedy resource allocator scales down the job with the longest training time, but this may result in computationally intensive training jobs having their node count reduced repeatedly, thereby resulting in unnecessarily long training times.

Third, the greedy allocate decisions are short-sighted. The allocator only deals with what is currently happening in the elastic training system, but has no consideration for the future. Because the system will face different computational demands in the future, it is necessary to look ahead and plan computational resource allocation accordingly.

\section{Optimal Resource Allocator}\label{section: optimal}

This paper proposes an optimal resource allocator that can overcome the above limitations of the greedy one. This section presents the methodology behind it. Please refer to Table \ref{table: notation} for all notations. We adopt a rolling-horizon approach to plan resource allocation for the future. The optimization problem is first formulated as a mixed-integer non-linear programming model. Then we linearize and simplify the model into a mixed-integer linear one to make it solvable. 

\begin{table}[h!]
    \caption{Notations}
    \label{table: notation}
    \centering
    \begin{tabular}{ p{0.5in} p{2.2in} } 
 \hline
 Sets &   \\ 
 \hline
 $I$ & set of deep learning jobs $i$ \\ 
 $T$ & set of time steps $t$ in a planning time horizon, i.e., $\{1,2,3, \cdots \}$ \\
 $K$ & set of legal number of nodes $k$, i.e., $\{1,2,4,8,16,32, \cdots \}$ \\
 \hline
 Parameters & \\ 
 \hline
 $d_{i}$ & remaining ETA or computation demand of job $i \in I$ \\
 $n_{i,min}$ & minimum number of nodes for job $i \in I$ \\
 $n_{i,max}$ & maximum number of nodes for job $i \in I$ \\
 $N$ & number of nodes in resource pool \\
 $f$ & elastic training frequency (unit: minute) \\
 $p$ & time step adjustment parameter  \\
 $M$ & the big-$M$, a sufficiently large number \\
 \hline
 Decision variables &  \\
 \hline
 $n_{i}^{t}$ & allocated number of nodes for job $i \in I$ at time step $t \in T$, integer \\
 $n_{i}$ & implemented allocate decision for job $i \in I$, integer \\
 $s_{i}^{t}$ & served computation demand for job $i \in I$ at time step $t \in T$ \\
 $y_{i}^{t}$ & whether job $i \in I$ has finished training at time step $t \in T$, binary\\
 $\delta_{i,k}^{t,-}$, $\delta_{i,k}^{t,+}$ & indicators for selecting $k \in K$ nodes for job $i \in I$ at time step $t \in T$, binary \\
 \hline
 \end{tabular}
\end{table}

\subsection{A Rolling-horizon Approach}

The optimal resource allocator adopts a rolling-horizon approach. This approach discretizes a planning time horizon $T$ into multiple time steps $t \in T = \{1,2,3,...\}$, and the length of each time step equals the elastic training frequency $f$, e.g., 5 minutes. Therefore, this approach can look ahead and plan decisions for the future accordingly. Consider a job’s remaining ETA as its computation demand. For each job $i \in I$ at time step $t \in T$, use $n_i^t$ nodes to serve $s_i^t$ out of job $i$'s demand $d_i$.

\begin{figure}[h]
\centerline{\includegraphics[width=3.2in,keepaspectratio]{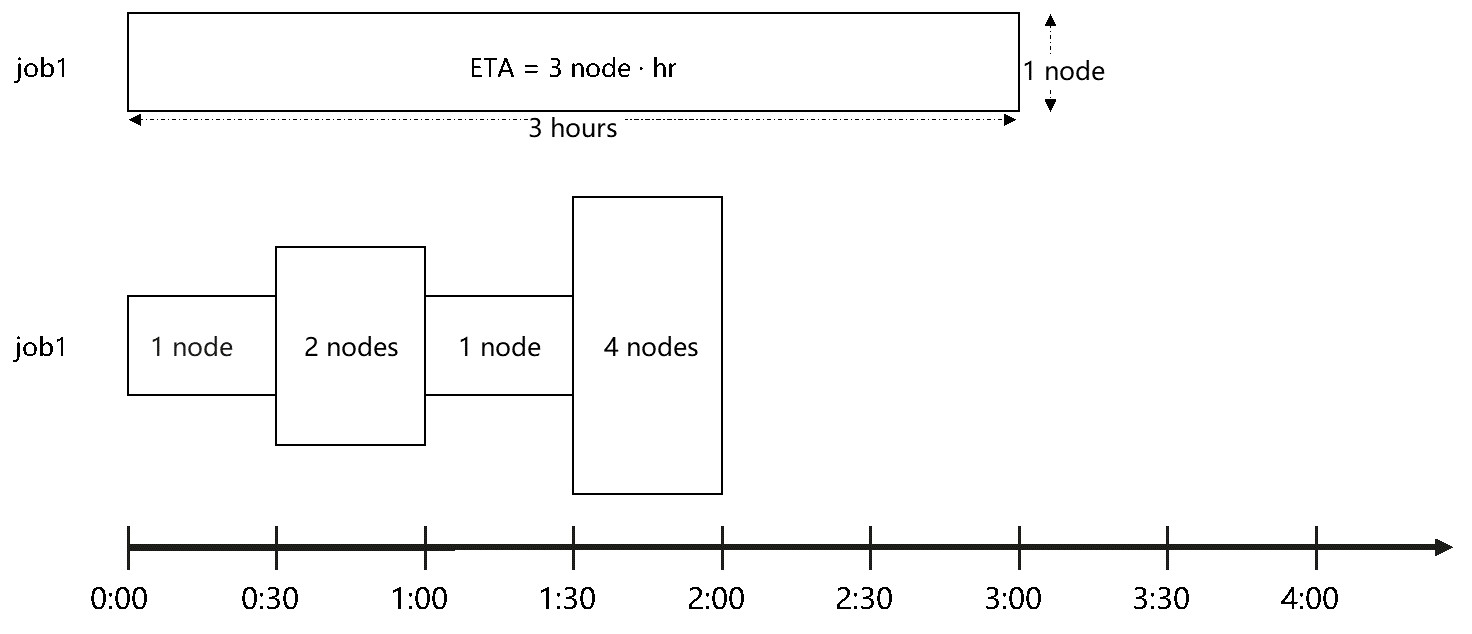}}
\caption{The rolling-horizon approach.}
\label{fig: rolling-horizon}
\end{figure}

Fig. \ref{fig: rolling-horizon} is an example in which the optimal resource allocator looks 4 hours ahead at 0:00. If elastic training frequency is 30 minutes, there will be 8 time steps. Job 1 is submitted at 0:00 and its ETA is 3 hours on a single node. We may use $n_1^1 = 1$ node to serve $s_1^1$ node$\cdot$hr demand out of the total 3 node$\cdot$hr. We may use 2, 1, and 4 nodes for the next three time steps, respectively, so job 1 might be done in only 2 hours.

\subsection{Non-linear Formulation}

The objective of optimal resource allocator is to maximize the training progress of all the jobs $I$ over the look-ahead time horizon $T$, as shown in (\ref{eq: obj}). A job's training progress is defined as its total served demand during $T$, i.e., $\sum_{t} s_i^t$, over its demand $d_i$.

\begin{equation}\label{eq: obj}
    \max \sum_{i} \sum_{t} \frac{s_{i}^{t}}{d_{i}} 
\end{equation}

subject to
\begin{equation}\label{eq: < demand}
    s_{i}^{t} \leq d_{i}
    \quad \forall i \in I, 
    \forall t \in T
\end{equation}
\begin{equation}\label{eq: training progress 1}
    s_{i}^{t=1} \leq p \cdot n_{i}^{t=1} \cdot 0.8^{\log _{2} n_{i}^{t=1}}
    \quad \forall i \in I
\end{equation}
\begin{equation}\label{eq: training progress 2}
    s_{i}^{t} \leq s_{i}^{t-1} + p \cdot n_{i}^{t} \cdot 0.8^{\log _{2} n_{i}^{t}}
    \quad \forall i \in I, 
    \forall t \in T, t \geq 2
\end{equation}
\begin{equation}\label{eq: < max}
    n_{i}^{t} \leq n_{i,max}
    \quad \forall i \in I, 
    \forall t \in T
\end{equation}
\begin{equation}\label{eq: < total nodes}
    \sum_{i} n_{i}^{t} \leq N
    \quad \forall t \in T
\end{equation}
\begin{equation}\label{eq: define y 1}
    y_{i}^{t} \leq \frac{s_{i}^{t}}{d_{i}} 
    \quad \forall i \in I, 
    \forall t \in T
\end{equation}
\begin{equation}\label{eq: define y 2}
    y_{i}^{t} \geq 1 - M(1 - \frac{s_{i}^{t}}{d_{i}})
    \quad \forall i \in I, 
    \forall t \in T
\end{equation}
\begin{equation}\label{eq: > min at first time step}
     n_{i}^{t=1} \geq n_{i,min}
    \quad \forall i \in I
\end{equation}
\begin{equation}\label{eq: > min if not finish}
     n_{i}^{t} \geq n_{i,min} - M \cdot y_{i}^{t-1}
    \quad \forall i \in I, 
    \forall t \in T,
    t \geq 2
\end{equation}
\begin{equation} \label{eq: return 0 if finish}
    n_{i}^{t} \leq M(1 - \frac{s_{i}^{t-1}}{d_{i}})
    \quad \forall i \in I, 
    \forall t \in T,
    t \geq 2
\end{equation}
\begin{equation}\label{eq: variable bound n}
    n_i^t \geq 0, integer \quad \forall i \in I, \forall t \in T
\end{equation}
\begin{equation}\label{eq: variable bound s}
    s_i^t \geq 0 \quad \forall i \in I, \forall t \in T
\end{equation}
\begin{equation}\label{eq: variable bound y}
    y_i^t \in \{0,1\} \quad \forall i \in I, \forall t \in T
\end{equation}

\begin{figure}[h]
\centerline{\includegraphics[width=3.2in,keepaspectratio]{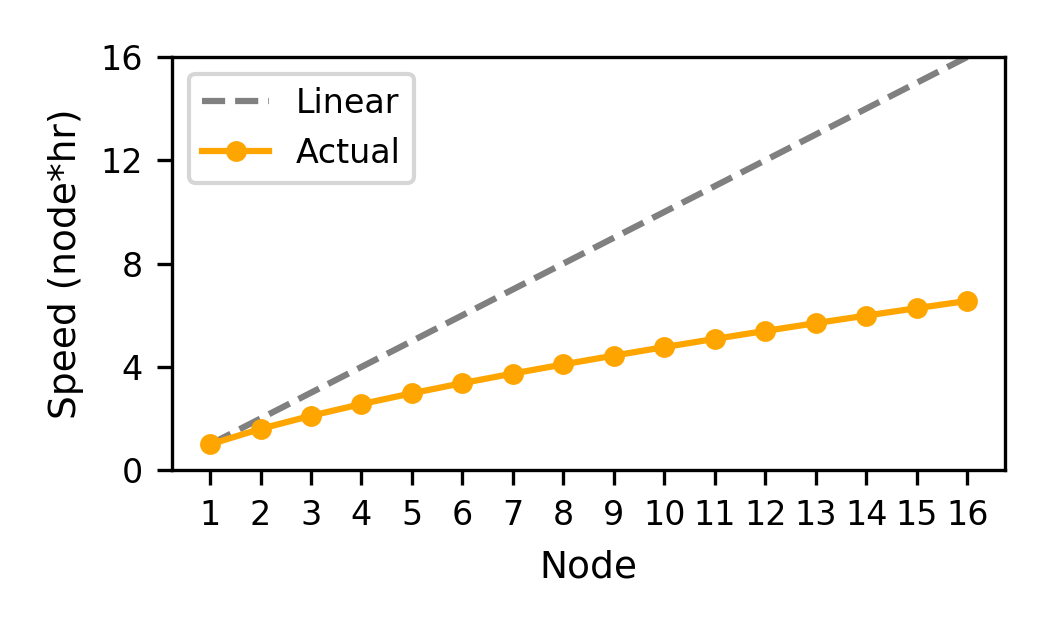}}
\caption{Relationship between number of nodes and training speed.}
\label{fig: node vs training speed}
\end{figure}

Constraint (\ref{eq: < demand}) makes sure that served demand $s_i^t$ at time step $t$ cannot overtake the job's demand. Constraint (\ref{eq: training progress 1}) and (\ref{eq: training progress 2}) are the training process in which $p$ is the time step adjustment parameter, for example, $p$ = 30/60 = 0.5, if each time step is 30 minutes. Given $n_i^t$ nodes to train job $i$ at time step $t$, the served demand will accumulate by $p \cdot n_{i}^{t} \cdot 0.8^{\log _{2} n_{i}^{t}}$. This non-linear relationship between number of nodes and training speed, as shown in Fig. \ref{fig: node vs training speed} is found from the historical data of Huawei ModelArts \cite{b3}. The ideal training speed is linearly related to the number of nodes, however, the actual training speed decreases by 20\% on average when the number of nodes doubles (i.e., multiply linear training speed by 0.8 every time the number of nodes doubles). While a job can be trained faster using more nodes, however, the speed attenuation becomes larger as node count increases. System-level training efficiency is likely to downgrade if only a few jobs are monopolizing all the resources. 

Constraint (\ref{eq: < max}) and (\ref{eq: < total nodes}) are straight-forward. Every job should have an upper bound of node size, denoted as $n_{i,max}$. The allocated nodes must not overtake total resources, i.e., $N$ nodes. However, the lower bound conditions are quite tricky. On the one hand, we must allocate a minimum number of nodes $n_{i,min}$ (usually 1 node) when job $i$ is training. On the other hand, if job $i$ has finished training at time step $t$, then its node size should return back to zero starting from the next time step $t+1$. To satisfy the two requirements, we have to introduce an extra binary variable $y_i^t$ to indicate whether job $i$ has finished training at time step $t$ and a sufficiently-large number $M$. If job $i$ has not finished yet, $y_i^t=0$; otherwise, $y_i^t=1$, as shown in Constraint (\ref{eq: define y 1}) and (\ref{eq: define y 2}). Constraint (\ref{eq: > min at first time step}) and (\ref{eq: > min if not finish}) ensure that minimum resource is allocated if a job is still training. Constraint (\ref{eq: return 0 if finish}) let node size return back to zero after a job has done. Constraints (\ref{eq: variable bound n})$\sim$(\ref{eq: variable bound y}) are the bounds of decision variables.

However, the above formulation is a mixed-integer non-linear programming model and hard to solve. The training process constraints (\ref{eq: training progress 1}) and (\ref{eq: training progress 2}) are non-linear and may be not easy to satisfy. Constraints (\ref{eq: define y 1})-(\ref{eq: return 0 if finish}) contain too many integer decision variables and constraints, which may make the model very slow to find optimal solutions. Note that decision-making for our resource allocation problem should be real-time or near real-time. In addition, the allocate number of nodes for each job must be $2^m$ ($m \in \mathbb{N}$) out of training accuracy reasons, i.e., $n_i^t \in K=\{1,2,4,8,16,32,...\}$, but this formulation cannot meet this requirement.

\subsection{Linearized and Simplified Formulation}

\begin{figure}[h]
\centerline{\includegraphics[width=3.2in,keepaspectratio]{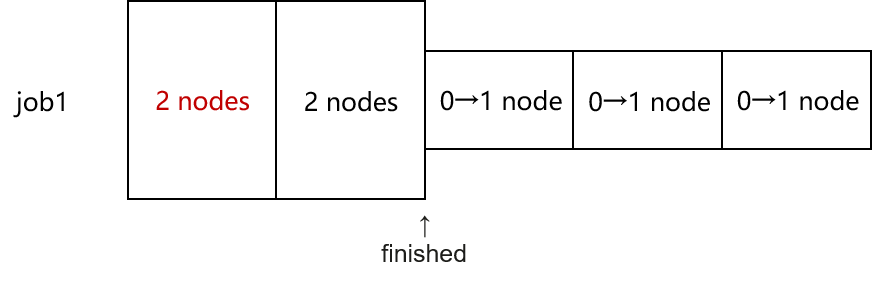}}
\caption{An example for nodes not returning back to zero after a job has finished.}
\label{fig: not return 0}
\end{figure}

We must linearize the non-linear constraints and simplify the above model formulation to make it solvable and fast. First, we drop the binary variables $y_i^t$ and replace constraints (\ref{eq: define y 1})-(\ref{eq: return 0 if finish}) with a new constraint (\ref{eq: > min}). This constraint allows nodes not return back to zero after a job has finished training. The impact is minimal because the optimal resource allocator adopts the rolling-horizon approach in which only the decision in the first time step, i.e., $n_i = n_i^{t=1}$, will be implemented. Taking Fig. \ref{fig: not return 0} as an example, after the job has finished training at the second time step, we allow the minimum number of nodes (1 node) in the optimal solution and allocate 2 nodes (the decision of the first time step) to train this job. The  problem size decreases significantly by simplifying model formulation. For a problem with 100 jobs and 5 time steps, we can save 500 integer variables and 1800 constraints.

\begin{equation}\label{eq: > min}
    n_{i}^{t} \geq n_{i,min}
    \quad \forall i \in I, 
    \forall t \in T
\end{equation}

The powers-of-two requirement can be met by introducing two binary indicators, $\delta_{i,k}^{t,-}$ and $\delta_{i,k}^{t,+}$, and constraints (\ref{eq: delta -})-(\ref{eq: exactly hold}). Allocate $n_i^t=k \in K$ nodes to job $i$ at time step $t$ if and only if $\delta_{i,k}^{t,-} = \delta_{i,k}^{t,+} = 1$. Constraints (\ref{eq: delta -}) and (\ref{eq: delta +}) add lower and upper bounds to the difference between $n_i^t$ and $k$. Constraint (\ref{eq: exactly hold}) ensures that exactly $|K|+1$ inequalities in (\ref{eq: delta -}) and (\ref{eq: delta +}) must hold, thus the value of $k$ will be a selection from the set $K$. For example, if the range of elastic training is from 1 node to 16 nodes, i.e., $k \in K=\{1,2,4,8,16\}$, we may allocate $n_1^1=k=4$ nodes to job 1 at time step 1 if and only if $\delta_{1,4}^{1,-} = \delta_{1,4}^{1,+} = 1$. In this case, exactly $|K|+1=6$ inequalities must hold, i.e., $n_1^1 \geq 1$, $n_1^1 \geq 2$, $n_1^1 \leq 4$, $n_1^1 \geq 4$, $n_1^1 \leq 8$, and $n_1^1 \leq 16$, while the remaining 4 inequalities, i.e., $n_1^1 \leq 1$, $n_1^1 \leq 2$, $n_1^1 \geq 8$, and $n_1^1 \geq 16$, must not hold. By this method, the optimal value of each $n_i^t$ must be $2^m$ ($m \in \mathbb{N}$).

\begin{equation}\label{eq: delta -}
\begin{split}
       \frac{1-\delta_{i,k}^{t,-}}{M} - M \cdot \delta_{i,k}^{t,-} \leq n_{i}^{t} - k \leq M \cdot (1 - \delta_{i,k}^{t,-}) \\
    \quad \forall i \in I, 
    \forall t \in T, \forall k \in K 
\end{split}
\end{equation}
\begin{equation}\label{eq: delta +}
\begin{split}
        \frac{1-\delta_{i,k}^{t,+}}{M} - M \cdot \delta_{i,k}^{t,+} \leq k - n_{i}^{t} \leq M \cdot (1 - \delta_{i,k}^{t,+}) \\ 
    \quad \forall i \in I, 
    \forall t \in T, \forall k \in K
\end{split}
\end{equation}
\begin{equation}\label{eq: exactly hold}
    \sum_{k} \delta_{i,k}^{t,-} + \sum_{k} \delta_{i,k}^{t,+} = |K|+1
    \quad \forall i \in I, 
    \forall t \in T
\end{equation}

Since binary variables $\delta_{i,k}^{t,-}$ and $\delta_{i,k}^{t,+}$ indicate whether the allocated number of nodes is a selection from the set $K$, we can replace constraints (\ref{eq: training progress 1})-(\ref{eq: training progress 2}) with two linear constraints (\ref{eq: linear 1})-(\ref{eq: linear 2}). If a $k \in K$ is selected, the training speed will exactly be $k \cdot 0.8^{\log _{2} k}$, as shown by the actual training speed curve in Fig. \ref{fig: node vs training speed}.

\begin{equation}\label{eq: linear 1}
\begin{split}
     s_{i}^{t=1} \leq p \cdot \sum_{k} k \cdot 0.8^{\log _{2} k} \cdot (\delta_{i,k}^{t=1,-} + \delta_{i,k}^{t=1,+} - 1) \\
    \quad \forall i \in I   
\end{split}
\end{equation}
\begin{equation}\label{eq: linear 2}
\begin{split}
       s_{i}^{t} \leq s_{i}^{t-1} + p \cdot \sum_{k} k \cdot 0.8^{\log _{2} k} \cdot (\delta_{i,k}^{t,-} + \delta_{i,k}^{t,+} - 1) \\
    \quad \forall i \in I, 
    \forall t \in T, t \geq 2 
\end{split}
\end{equation}

\textbf{\textit{Proposition}}. The binary variables $\delta_{i,k}^{t,-}$ and $\delta_{i,k}^{t,+}$ do not introduce estimation errors to actual training speeds.

\textbf{\textit{Proof}}. If $k^{*} \in K$ is selected as the optimal solution for job $i$ at time step $t$, then we have $\delta_{i,k^{*}}^{t,-}=\delta_{i,k^{*}}^{t,+}=1$. By (\ref{eq: exactly hold}), it is easy to know that for $k \in K$ and $k \neq k^{*}$, $\delta_{i,k}^{t,-}+\delta_{i,k}^{t,+}=1$ must hold. Therefore, the training speed is $\sum_{k} k \cdot 0.8^{\log _{2} k} \cdot (\delta_{i,k}^{t,-} + \delta_{i,k}^{t,+} - 1) = 1 \cdot 0.8^{\log _{2} 1} \cdot (1 - 1) + \cdots + k^{*} \cdot 0.8^{\log _{2} k^{*}} \cdot (2 - 1) + \cdots = k^{*} \cdot 0.8^{\log _{2} k^{*}}$, which is exactly the actual speed as shown in Fig. \ref{fig: node vs training speed}.

\subsection{Final Formulation}

In summary, we re-formulate the optimal resource allocator as a mixed-integer linear program (MILP) as follows. We use the open-source optimization solver \textit{CBC} \cite{b11} to search for optimal solutions.

\begin{equation*}
    \max_{\textbf{n}} \quad \sum_{i} \sum_{t} \frac{s_{i}^{t}}{d_{i}} 
\end{equation*}
subject to
\begin{equation*}
    s_{i}^{t} \leq d_{i}
    \quad \forall i \in I, 
    \forall t \in T
\end{equation*}
\begin{equation*}
    n_{i}^{t} \leq n_{i,max}
    \quad \forall i \in I, 
    \forall t \in T
\end{equation*}
\begin{equation*}
    \sum_{i} n_{i}^{t} \leq N
    \quad \forall t \in T
\end{equation*}
\begin{equation*}
    n_{i}^{t} \geq n_{i,min}
    \quad \forall i \in I, 
    \forall t \in T
\end{equation*}
\begin{equation*}
\begin{split}
       \frac{1-\delta_{i,k}^{t,-}}{M} - M \cdot \delta_{i,k}^{t,-} \leq n_{i}^{t} - k \leq M \cdot (1 - \delta_{i,k}^{t,-}) \\
    \quad \forall i \in I, 
    \forall t \in T, \forall k \in K 
\end{split}
\end{equation*}
\begin{equation*}
\begin{split}
        \frac{1-\delta_{i,k}^{t,+}}{M} - M \cdot \delta_{i,k}^{t,+} \leq k - n_{i}^{t} \leq M \cdot (1 - \delta_{i,k}^{t,+}) \\ 
    \quad \forall i \in I, 
    \forall t \in T, \forall k \in K
\end{split}
\end{equation*}
\begin{equation*}
    \sum_{k} \delta_{i,k}^{t,-} + \sum_{k} \delta_{i,k}^{t,+} = |K|+1
    \quad \forall i \in I, 
    \forall t \in T
\end{equation*}
\begin{equation*}
\begin{split}
     s_{i}^{t=1} \leq p \cdot \sum_{k} k \cdot 0.8^{\log _{2} k} \cdot (\delta_{i,k}^{t=1,-} + \delta_{i,k}^{t=1,+} - 1) \\
    \quad \forall i \in I   
\end{split}
\end{equation*}
\begin{equation*}
\begin{split}
       s_{i}^{t} \leq s_{i}^{t-1} + p \cdot \sum_{k} k \cdot 0.8^{\log _{2} k} \cdot (\delta_{i,k}^{t,-} + \delta_{i,k}^{t,+} - 1) \\
    \quad \forall i \in I, 
    \forall t \in T, t \geq 2 
\end{split}
\end{equation*}
\begin{equation*}
    s_i^t \geq 0 \quad \forall i \in I, \forall t \in T
\end{equation*}
\begin{equation*}
    n_i^t \geq 0, integer \quad \forall i \in I, \forall t \in T
\end{equation*}
\begin{equation*}
    \delta_{i,k}^{t,-}, \delta_{i,k}^{t,+} \in \{0,1\} \quad \forall i \in I, \forall t \in T, \forall k \in K
\end{equation*}

\section{Simulation}\label{section: simulation}

To compare the proposed optimal resource allocator with the greedy one that Huawei Cloud previously implemented, this section designs a system that simulates the training process of deep learning jobs on the cloud. Several experiments were conducted to examine each resource allocator's queueing delay, training efficiency, performance on heterogeneous-ETA jobs, robustness to ETA disturbance, and performance under scaling delay. In addition, this section shows efficiency of the optimal resource allocator.

\subsection{Simulation Setup}

\begin{figure}[h]
\centerline{\includegraphics[width=3.2in,keepaspectratio]{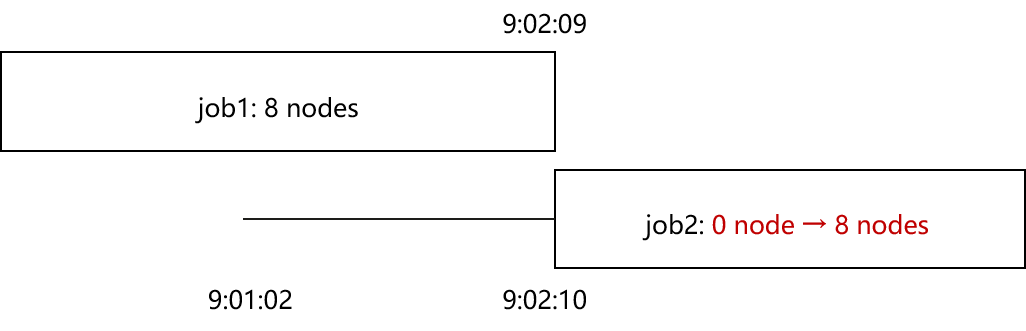}}
\caption{Example for resource allocation at regular times in simulation.}
\label{fig: allocate max}
\end{figure}

We set up a simulation framework to conduct experiments for the elastic training system. The system’s allocate decisions are made either by the optimal resource allocator or the greedy resource allocator. Simulation moves forward second by second. For each second, the system could keep training some jobs, initiate queueing jobs, or finish training some others. Simulation updates the status of each job each second.
Resource allocator makes decisions every 5 minutes, for example, 9:00:00, 9:05:00 and 9:10:00. The look-ahead time horizon is 25 minutes and the length of each time step equals the elastic training frequency, so there are 5 time steps. When simulation arrives at scaling moments, resource allocator activates and tells resource pool how to scale up or down jobs. During regular times, simulation allocates maximum resources to queueing jobs, if any. For example, in Fig. \ref{fig: allocate max} job 1 finishes training at 9:02:09 and releases 8 nodes. Job 2 joins queue since 9:01:02, so 1 second later the system allocates the 8 nodes to it.

Data used for simulation come from Huawei ModelArts for deep learning jobs. There were 1252 jobs submitted from January 24, 2021 14:25:41 to January 26, 2021 16:07:19, spanning around 2 days and 2 hours. On average, a job spent 2.8 minutes in the queue, 84.3 minutes for training, and 87.1 minutes in total. Among these jobs 447 of them spent training time of $\geq$5 minutes. The average queueing time is 7.2 minutes, training time is 232.6 minutes, and total time is 239.9 minutes. We select the 447 jobs as simulation baseline.

\subsection{Queueing Delay and Training Efficiency}

\begin{figure}[h]
\centerline{\includegraphics[width=3.2in,keepaspectratio]{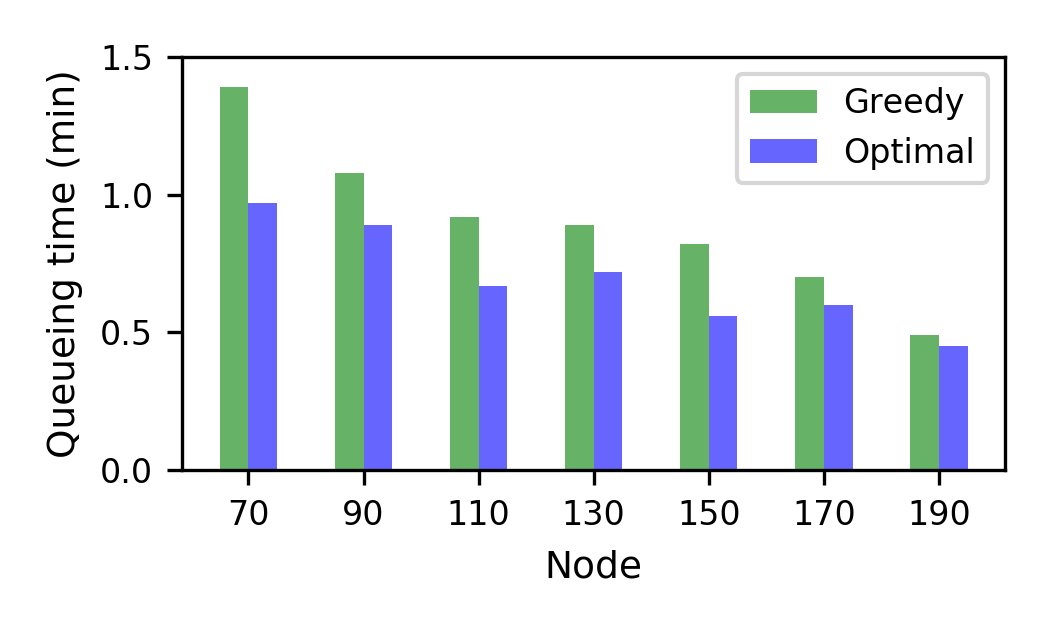}}
\caption{Comparison of queueing time.}
\label{fig: queueing delay}
\end{figure}

We assume that total computational resources are ranging from 70 to 190 nodes with an interval of 20 for the baseline data and examine the average queueing time per job. In Fig. \ref{fig: queueing delay}, queueing delay shows a downward trend as more nodes are available in resource pool for both greedy and optimal resource allocators. The optimal one always reduces queueing time given the same resources. The largest decrease is 32\% at 150 nodes.

\begin{figure}[h]
\centerline{\includegraphics[width=3.2in,keepaspectratio]{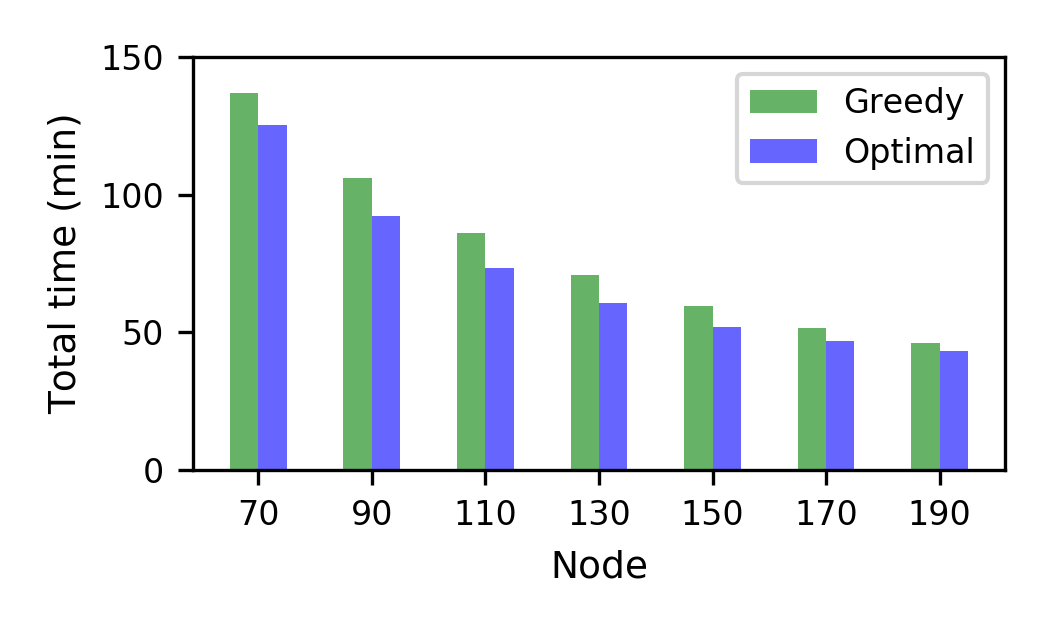}}
\caption{Comparison of total time.}
\label{fig: training efficiency 1}
\end{figure}

\begin{figure}[h]
\centerline{\includegraphics[width=3.2in,keepaspectratio]{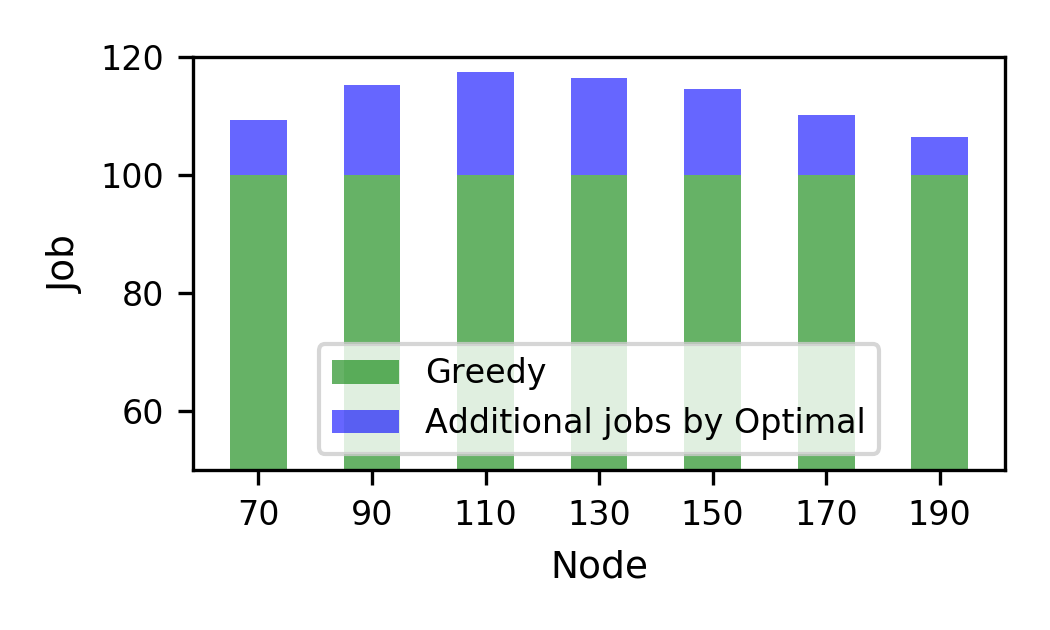}}
\caption{Additional jobs trained by optimal resource allocator.}
\label{fig: training efficiency 2}
\end{figure}

Total time is queueing time plus training time, and it indicates how long users can get training results after they have submitted a job. It is seen from Fig. \ref{fig: training efficiency 1} that total time is declining as more available nodes train jobs for both resource allocators. The optimal one can always accelerate training given the same resources. The improvement is shown in Fig. \ref{fig: training efficiency 2}. When the greedy resource allocator has trained 100 jobs, the optimal one can train more and the additional trained jobs are illustrated by the blue bars. The improvement is up to 17.4 jobs given 110 nodes.

Note that efficiency improvement is bell-shaped. When computational resources are relatively limited, jobs tend to occupy as fewer nodes as possible, no matter what resource allocator is. An extreme example could be that the number of training jobs equals node size, so the optimal decision would be to let each job occupy only one node. On the contrary, when computational resources are abundant the best decision would be to allocate as many nodes as possible to all jobs. Thus, the gap between the two resource allocators becomes narrower.

\subsection{Impact of Heterogeneous-ETA Jobs}

Small-ETA jobs with training time of less than 5 minutes account for 64\% in our simulation dataset. Elastically training small-ETA jobs or not remains a question. On the one hand, these jobs might take a little time to finish training even if the allocated nodes are little. On the other hand, small-ETA jobs take a significant portion and may slow down the entire training efficiency if their resources are insufficient.

\begin{figure}[h]
\centerline{\includegraphics[width=3.2in,keepaspectratio]{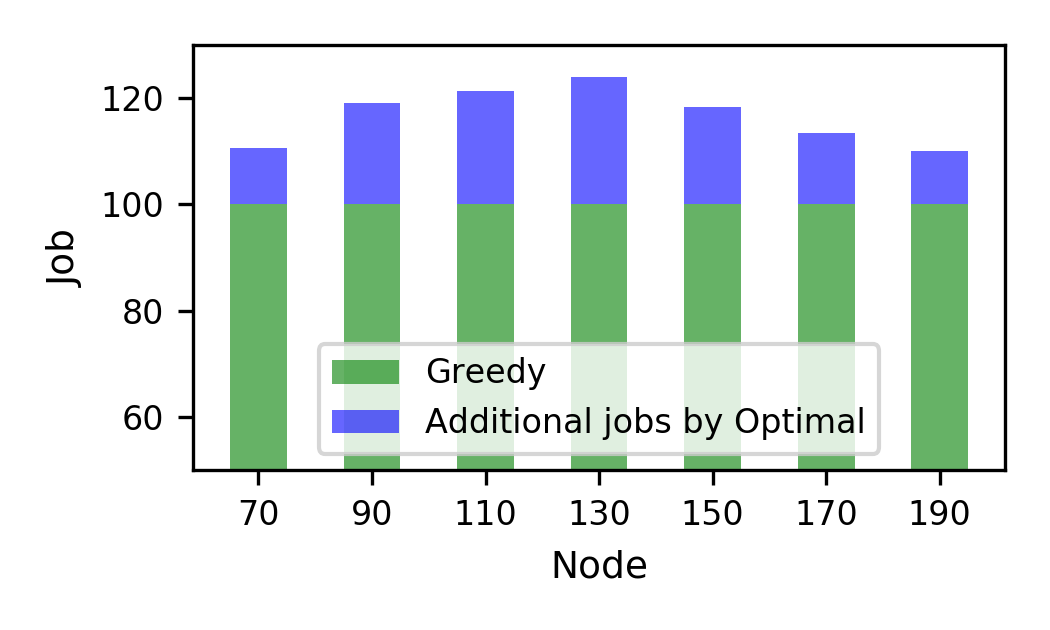}}
\caption{Additional trained jobs by optimal resource allocator when ETA becomes more heterogeneous.}
\label{fig: small eta}
\end{figure}

For this simulation, we let elastic training system scales all 1252 jobs whatever their ETA is. Therefore, job ETA becomes more heterogeneous than the baseline scenario. Fig. \ref{fig: small eta} shows that up to additional 24.1 jobs can be completed by optimal resource allocator, compared to the baseline 17.4 jobs in Fig. \ref{fig: training efficiency 2}. Therefore, the optimal resource allocator can better handle heterogeneous-ETA jobs than the greedy one.

\subsection{Robustness}

\begin{figure}[h]
\centerline{\includegraphics[width=3.2in,keepaspectratio]{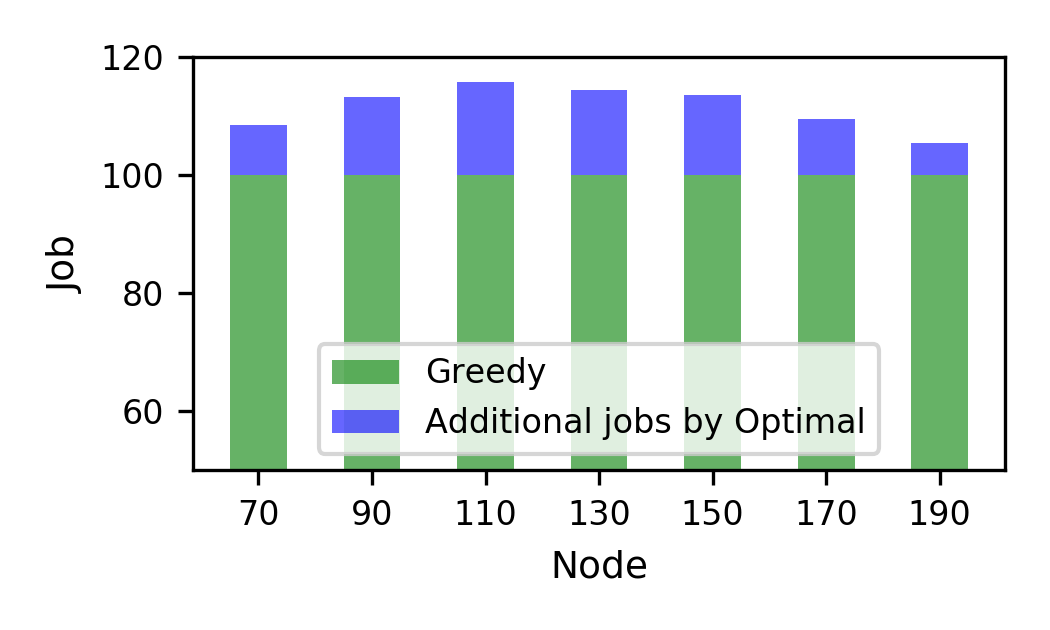}}
\caption{Robustness under ±10\% ETA disturbance.}
\label{fig: robust 10}
\end{figure}

It is almost impossible to predict job ETA 100\% accurately \cite{b12, b13, b14}. We add ±10\% disturbance to ETA for the baseline scenario in simulation. For example, if a job's actual runtime is 10 node$\cdot$hr, its ETA will be a random sample from the range of 9$\sim$11  node$\cdot$hr. Fig. \ref{fig: robust 10} shows the additional trained jobs by optimal resource allocator with the disturbance. Compared to the no-disturbance baseline in Fig. \ref{fig: training efficiency 2}, the additional trained jobs are only 0.6$\sim$2.4 fewer.

\begin{figure}[h]
\centerline{\includegraphics[width=3.2in,keepaspectratio]{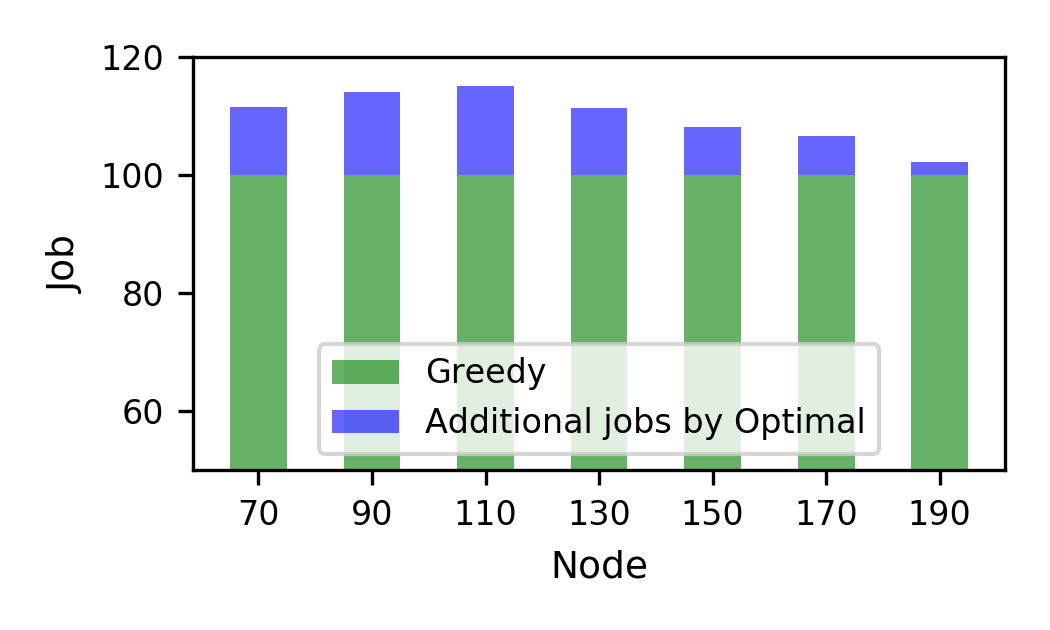}}
\caption{Robustness under ETA disturbance, bug jobs and user-terminated jobs.}
\label{fig: robust bug}
\end{figure}

It is also common that users submit jobs that contain bugs. Bug jobs usually hang up within a few minutes once starting training. Their ETA and actual training duration could be significantly different because bugs are almost unpredictable. Users may also terminate jobs that are training at any time out of many reasons, e.g., losing patience. We conduct a simulation based on the baseline data in which 75\% jobs have ±10\% ETA disturbance, 15\% jobs contain bugs and hang up randomly within 5 minutes, and 10\% jobs could be terminated by users at any time. Fig. \ref{fig: robust bug} shows that there are still up to 15.0 additional jobs completed by optimal resource allocator under such harsh conditions.

The reasons optimal resource allocator is robust are that it adopts a rolling-horizon approach and makes allocate decisions every a few minutes. Even if estimation errors, bug jobs and user-terminated jobs enlarge the difference between ETA and actual runtime, the negative impact on system training efficiency is small.

\subsection{Impact of Scaling Delay}

\begin{figure}[h]
\centerline{\includegraphics[width=3.2in,keepaspectratio]{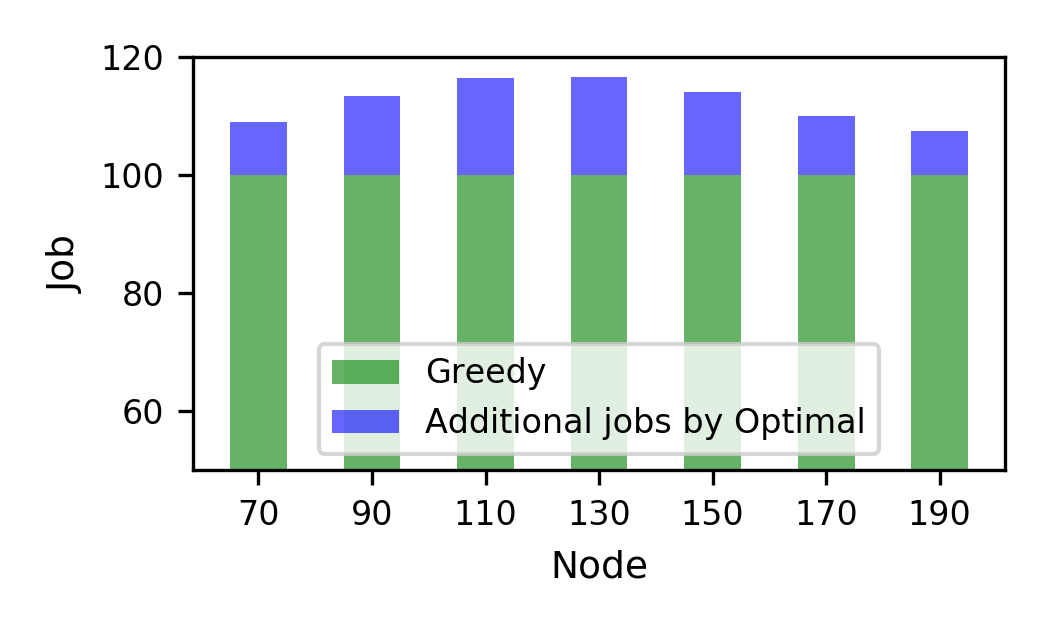}}
\caption{Additional trained jobs by optimal resource allocator with 15-sec scaling delay.}
\label{fig: scale delay}
\end{figure}

The above simulations assume no delay when a job initiates or scales. The time of scaling-down is usually minimal while initiation or scaling-up takes 10$\sim$20 seconds on average. For this simulation, we take 15-second delay into consideration. When a job just starts training or resource allocator decides to scale up a job, it will stay on its current nodes for 15 seconds before executing the new decision. Fig. \ref{fig: scale delay} shows the additional trained jobs by optimal resource allocator with 15-second delay. The differences compared to the baseline scenario are merely -0.9$\sim$1.8 additional jobs. The reason for the differences could be that during the 15-second delay some jobs are still executing the last allocate decisions that are not optimal at this moment, so the additional trained jobs become slightly fewer.

\subsection{Speed of Optimal Resource Allocator}

\begin{figure}[h]
\centerline{\includegraphics[width=3.2in,keepaspectratio]{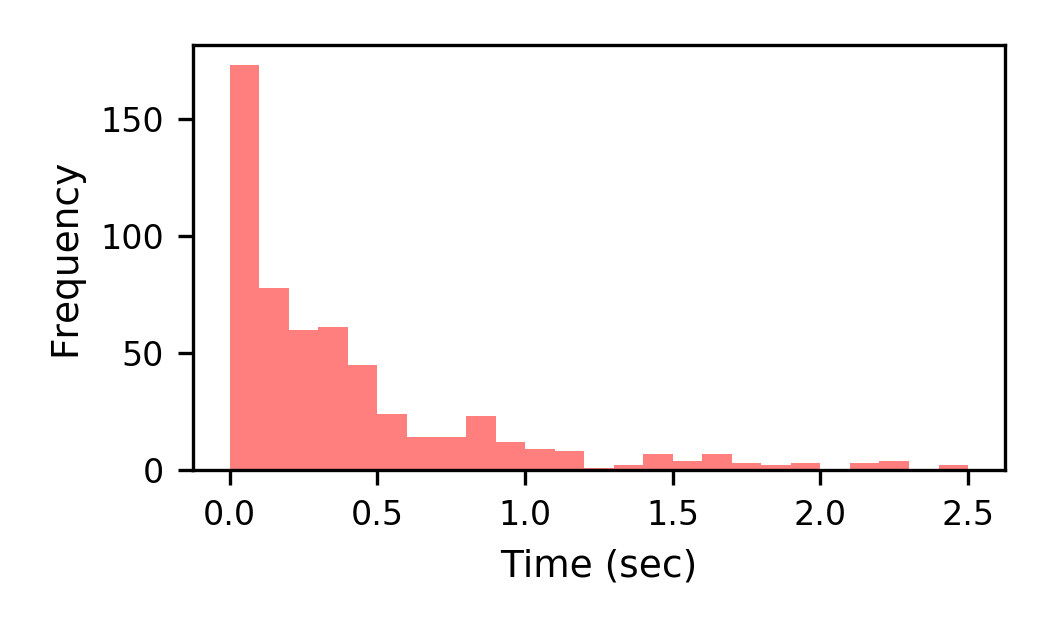}}
\caption{Histogram of solution time by optimal resource allocator.}
\label{fig: speed}
\end{figure}

Resource allocation for elastic training should be real-time or near real-time. We have recorded the time spent for every decision-making by optimal resource allocator in simulation. The baseline scenario is provided with 70 nodes. The simulation runs on a Linux server with Intel i7-8700K CPU and 64 GB RAM. Fig. \ref{fig: speed} is the histogram of solution time and shows the optimal resource allocation is very fast in decision-making, spending merely 0.4 seconds on average. The median time is 0.24 seconds. For 95\% cases, solution time is under 1.49 seconds (95\textit{p}\textsuperscript{th}), and the maximum time is not over 2.48 seconds.

\section{Conclusion}\label{section: conclusion}

This paper proposes an optimal resource allocator of elastic training for deep learning jobs on cloud. The allocator adopts a rolling-horizon approach and maximizes training progress of all jobs over a planning time horizon. The original model formulation contains non-linear constraints and many integer variables, which are hard to optimize. We simplify and linearize the model formulation into a MILP that is smaller-scaled, more solvable and faster. We also introduce an innovative method to make sure that allocated number of nodes meet the powers-of-two requirements. 

We design a simulation framework to conduct experiments to elastic training systems with the optimal resource allocator and a greedy one as benchmark. For the baseline scenario, the optimal resource allocator can reduce  queueing time by up to 32\% and accelerate training efficiency by up to 17.4\%, which greatly improves user experience. Also, the optimal resource allocator can better handle heterogeneous-ETA jobs than the greedy one, training up to 24.1 additional jobs. Simulations that test robustness show that the optimal allocator is very robust to ETA disturbance, bug jobs and user-terminated jobs. The impact of 15-second scaling delay is also examined and the additional trained jobs are merely different from the no-delay baseline. Also, searching for optimal solutions is very fast, taking only 0.4 seconds on average.

\bibliographystyle{abbrv}

\begin{thebibliography}{00}
\bibitem{b1} Amazon Web Services. Getting Started with AWS. Retrieved from https://aws.amazon.com/getting-started/.
\bibitem{b2} Microsoft Azure. Get to know Azure. Retrieved from https://azure.microsoft.com/en-ca/overview/.
\bibitem{b3} Huawei Cloud. ModelArts. Retrieved from https://www.huaweicloud.com/intl/en-us/product/modelarts.html.
\bibitem{b4} Alibaba Cloud. Overview of Auto Scaling. Retrieved from https://bit.ly/3aBlIWi.
\bibitem{b5} Lin, H., Zhang, H., Ma, Y., He, T., Zhang, Z., Zha, S. and Li, M., 2019. Dynamic mini-batch SGD for elastic distributed training: learning in the limbo of resources. arXiv preprint arXiv:1904.12043.
\bibitem{b6} Kubernetes. Production-Grade Container Orchestration. Retrieved from https://kubernetes.io/.
\bibitem{b7} Amazon Web Services. Amazon Elastic Container Service. Retrieved from https://aws.amazon.com/ecs/.
\bibitem{b8} Red Hat. Red Hat OpenShift Container Platform. Retrieved from https://www.openshift.com/products/container-platform.
\bibitem{b8.1} Shen, Z., Subbiah, S., Gu, X. and Wilkes, J., 2011, October. Cloudscale: elastic resource scaling for multi-tenant cloud systems. In Proceedings of the 2nd ACM Symposium on Cloud Computing (pp. 1-14).
\bibitem{b8.2} Gregory, A. and Majumdar, S., 2016, March. A constraint programming based energy aware resource management middleware for clouds processing mapreduce jobs with deadlines. In Companion Publication for ACM/SPEC on International Conference on Performance Engineering (pp. 15-20).
\bibitem{b8.3} Liu, L. and Xu, H., 2018, October. Elasecutor: Elastic executor scheduling in data analytics systems. In Proceedings of the ACM Symposium on Cloud Computing (pp. 107-120).
\bibitem{b8.4} Javadi, S.A., Suresh, A., Wajahat, M. and Gandhi, A., 2019, November. Scavenger: A black-box batch workload resource manager for improving utilization in cloud environments. In Proceedings of the ACM Symposium on Cloud Computing (pp. 272-285).
\bibitem{b8.5} Chen, Y., Peng, Y., Bao, Y., Wu, C., Zhu, Y. and Guo, C., 2020, October. Elastic parameter server load distribution in deep learning clusters. In Proceedings of the 11th ACM Symposium on Cloud Computing (pp. 507-521).
\bibitem{b8.6} Chen, C., Weng, Q., Wang, W., Li, B. and Li, B., 2020, October. Semi-dynamic load balancing: efficient distributed learning in non-dedicated environments. In Proceedings of the 11th ACM Symposium on Cloud Computing (pp. 431-446).
\bibitem{b9} Saxena, V., Jayaram, K.R., Basu, S., Sabharwal, Y. and Verma, A., 2020, November. Effective elastic scaling of deep learning workloads. In 2020 28th International Symposium on Modeling, Analysis, and Simulation of Computer and Telecommunication Systems (MASCOTS) (pp. 1-8). IEEE.
\bibitem{b10} Zhu, J., Huang, Z., Wu, R., Bai, X., Yang, B., Li, Y., Zheng, H. and Dai, Z. 2020. A Design and Implementation Method for Elastic Distributed Training Systems. Chinese patent, 87068967CN02.
\bibitem{b11} Forrest, J. and Lougee-Heimer, R. CBC User Guide. Retrieved from https://www.coin-or.org/Cbc/cbcuserguide.html.
\bibitem{b12} Pham, T.P., Durillo, J.J. and Fahringer, T., 2017. Predicting workflow task execution time in the cloud using a two-stage machine learning approach. IEEE Transactions on Cloud Computing, 8(1), pp.256-268.
\bibitem{b13} Sidhanta, S., Golab, W. and Mukhopadhyay, S., 2016, May. Optex: A deadline-aware cost optimization model for spark. In 2016 16th IEEE/ACM International Symposium on Cluster, Cloud and Grid Computing (CCGrid) (pp. 193-202). IEEE.
\bibitem{b14} Mustafa, S., Elghandour, I. and Ismail, M.A., 2018. A machine learning approach for predicting execution time of spark jobs. Alexandria engineering journal, 57(4), pp.3767-3778.
\end{thebibliography}

\end{document}